# MAGNETIC STRUCTURES IN A MODEL NICKELATE: MONTE CARLO STUDY

**V. S. Ryumshin[1,*], Y. D. Panov[1], A. S. Moskvin[1,2]**

[1]*Institute of Natural Sciences and Mathematics, Ural Federal University, 620002, Ekaterinburg, Russia*

[2]*Institute of Metal Physics, UD RAS 620180, Ekaterinburg, Russia*

**e-mail: vitaliy.riumshin@urfu.ru*

**Abstract**.

Within the framework of the classical Monte Carlo method with kinematic account of constant bosons concentration a model nickelate was studied, its magnetic structures for various values of exchange integrals considering up to third-nearest neighbors were determined.

## INTRODUCTION

Orthorhombic $RNiO_3$ exhibits a first-order metal-insulator phase transition to a charge-ordered insulating state, which is also accompanied by a structural phase transition [1,2]. In addition to the discussion of the metal-insulator transition in nickelates, special attention is paid to the magnetic structure, since at low temperatures, nickelates undergo a phase transition to an antiferromagnetic (AFM) state [2,3]. $PrNiO_3$ and $NdNiO_3$ are particularly notable, in which these three transitions occur simultaneously, which is the subject of active discussions [2,4,5]. There is still no consensus on what kind of magnetic structure is formed in the monoclinic phase of nickelates. Almost all studies note that this structure has an AFM nature with a propagation vector $\vec{k}$ = (1/2, 0, 1/2) or $\vec{k}$ = (1/4, 1/4, 1/4) in pseudocubic notation, however, experimental data allow for several variants of the magnetic structure due to the powder nature of the samples and different experimental techniques [1,6-8]. In particular, two structures are distinguished: collinear (such as "up-up-down-down") [3] and non-collinear (for example, a spiral order) [8]. In addition, for some representatives of the series, the possible presence of hidden phases [9] and the manifestation of magnetoelectric effects [10] are noted. A unified theory is also lacking. Most theoretical work is devoted to first-principles calculations, which identify the FM phase formed by nearest-neighbor interactions as energetically favorable [11,12]. However, these results do not take into account charge disproportionation, which leads to the suppression of the magnetic interaction between nearest neighbors due to the formation of charge ordering.

Previously, we investigated a model nickelate using the mean-field approximation and Monte Carlo methods [13–15]. These studies focused primarily on phase states, phase separation, metal-insulator transition and were limited to examining nearest-neighbor interactions. The goal of this study is to determine the magnetic structures of a model nickelate, taking into account the interactions of second and third neighbors, which are believed to be responsible for the formation of the magnetic structure in nickelates.

## MODEL AND METHODS

The model nickelate is a charge-disproportionated phase of $RNiO_3$, the electronic and magnetic structure of which is described using a generalized model of effective charge triplets [16,17]. The basis of states at a site includes 8 states: a nonmagnetic singlet $Ni^{4+}$, a Jahn-Teller quartet $Ni^{3+}$, and a spin triplet $Ni^{2+}$ (table 1).

**Table 1.** Pseudospin, spin and orbital structure of the $NiO_6$ centers in orthonickelates.

| $\hat{\Sigma}_z = M$ | Ion | Cluster | Configuration | Spin S | Degeneracy |
|---|---|---|---|---|---|
| 1 | $Ni^{4+}$ ($d^8$) | $[NiO_6]^{8-}$ | $t_{2g}^6$; $^1A_{1g}$ | 0 | 1 |
| 0 | $Ni^{3+}$ ($d^7$) | $[NiO_6]^{9-}$ | $t_{2g}^6 e_g$; $^2E_g$ | 1/2 | 4 |
| -1 | $Ni^{2+}$ ($d^6$) | $[NiO_6]^{10-}$ | $t_{2g}^6 e_g^2$; $^3A_{2g}$ | 1 | 3 |

The Hamiltonian of the model nickelate can be written as follows:

$$\hat{\mathcal{H}} = -t\sum_{\langle ij\rangle\sigma}\left(\hat{B}_i^{\sigma+}\hat{B}_j^{\sigma} + \hat{B}_j^{\sigma+}\hat{B}_i^{\sigma}\right) + \frac{U}{2}\sum_{\langle ij\rangle}\hat{\Sigma}_{zi}^2 + V\sum_{\langle ij\rangle}\hat{\Sigma}_{zi}\hat{\Sigma}_{zj} +$$

$$J_1\sum_{\langle ij\rangle}\left(\hat{\boldsymbol{s}}_i, \hat{\boldsymbol{s}}_j\right) + J_2\sum_{\langle ik\rangle}\left(\hat{\boldsymbol{s}}_i, \hat{\boldsymbol{s}}_k\right) + J_3\sum_{\langle il\rangle}\left(\hat{\boldsymbol{s}}_i, \hat{\boldsymbol{s}}_l\right), \quad (1)$$

where $t$ is the pair (spin-triplet boson) transfer integral, $U$ and $V$ are the local and nonlocal charge correlation parameters, $J_1$, $J_2$, $J_3$ are the exchange integrals for 1st, 2nd and 3rd neighbors, respectively; $\hat{\Sigma}_z$ is the pseudospin operator, $\hat{\boldsymbol{s}}_l$ is the spin operator S = 1, $\hat{B}^{\sigma+}$ and $\hat{B}^{\sigma}$ are the spin-triplet boson creation and annihilation operators.

The simulation were performed using the classical Monte Carlo method with the Metropolis algorithm and a modification to kinematically account for the condition of constant boson concentration, which allows us to simulate the evolution of the system with conservation of the average charge. We previously performed this modification for a quartet of states of model nickelate [14], and in this study the algorithm was adapted for an octet of states. For an octet of states, the wave function at a node follows:

$$|\Psi_i\rangle = c_{10}^i|10\rangle_i + c_{0\frac{1}{2}}^i\left|0\,\frac{1}{2}\right\rangle_i + c_{0-\frac{1}{2}}^i\left|0-\frac{1}{2}\right\rangle_i + c_{0\frac{1}{2}}^i\left|0\,\frac{1}{2}\right\rangle_i + c_{0-\frac{1}{2}}^i\left|0-\frac{1}{2}\right\rangle_i +$$

$$c_{-11}^i|-11\rangle_i + c_{-10}^i|-10\rangle_i + c_{-1-1}^i|1-1\rangle_i, \ (2)$$

where states are denoted as $|m\sigma\rangle$: $m$ – pseudospin projection, σ – spin projection; $c_{m\sigma}^i = r_{m\sigma}^i e^{i\varphi_{m\sigma}^i}$, $\varphi_{m\sigma}^i \in [0, 2\pi]$.

Parameterization for the wave function coefficients:

$$r_{10} = R\cos\theta, \ (3)$$

$$r_{0\frac{1}{2}} = R\sin\theta\cos\psi, \ (4)$$

$$r_{0-\frac{1}{2}} = R\sin\theta\sin\psi\cos\phi, \ (5)$$

$$r_{0\frac{1}{2}} = R\sin\theta\sin\psi\sin\phi\cos\alpha, \ (6)$$

$$r_{0-\frac{1}{2}} = R\sin\theta\sin\psi\sin\phi\sin\alpha\cos\beta, \ (7)$$

$$r_{-11} = R\sin\theta\sin\psi\sin\phi\sin\alpha\sin\beta\, cos\,\gamma, \ (8)$$

$$r_{-10} = R \sin\theta \sin\psi \sin\phi \sin\alpha \sin\beta \sin\gamma \cos\lambda, \quad (9)$$

$$r_{1-1} = R \sin\theta \sin\psi \sin\phi \sin\alpha \sin\beta \sin\gamma \sin\lambda. \quad (10)$$

To obtain a uniform sample in the space of states at a node, it is necessary to generate 15 random numbers uniformly distributed over the interval [0,1]. The standard Metropolis algorithm is used to change the system states; the algorithm for selecting states on a pair of nodes is follows:

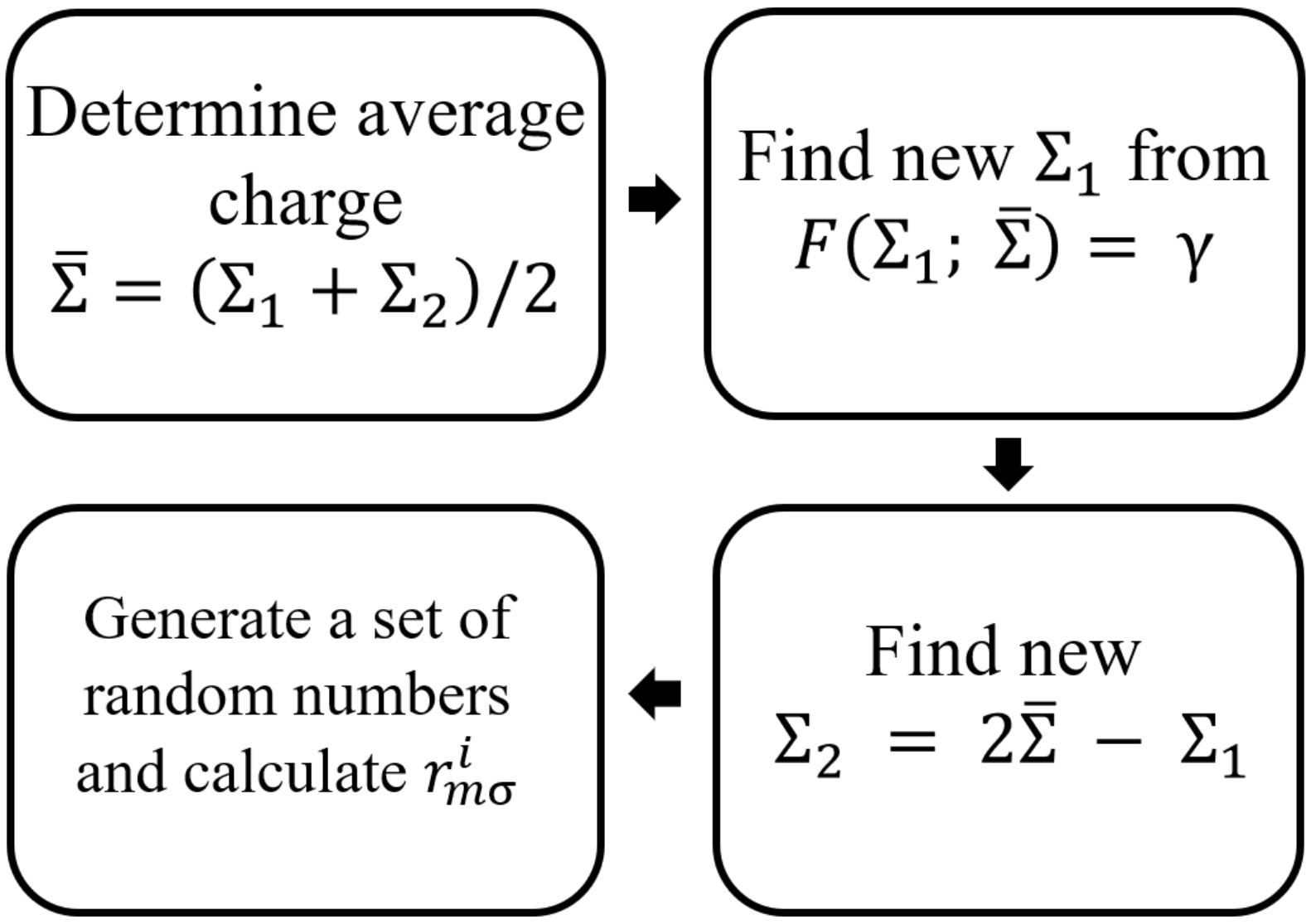

**Figure 1.** Algorithm for selecting states on a pair of nodes.

Magnetic structures are determined by calculating the magnetic structure factor $S(\vec{q})$ (11), which was implemented using the cuFFT library.

$$S(\vec{q}) = \left|\frac{1}{N}\sum_i e^{i\vec{q}\vec{r}_i}\langle\hat{\boldsymbol{s}}_i\rangle\right|^2. \quad (11)$$

The calculation program was implemented using CUDA technology to enable parallel calculations. The simulation was performed for a system of size 96x96 with $10^6$ Monte Carlo steps for annealing and measurements.

## RESULTS

Figure 1 shows the results of the structure factor for model nickelate at T = 0.001.

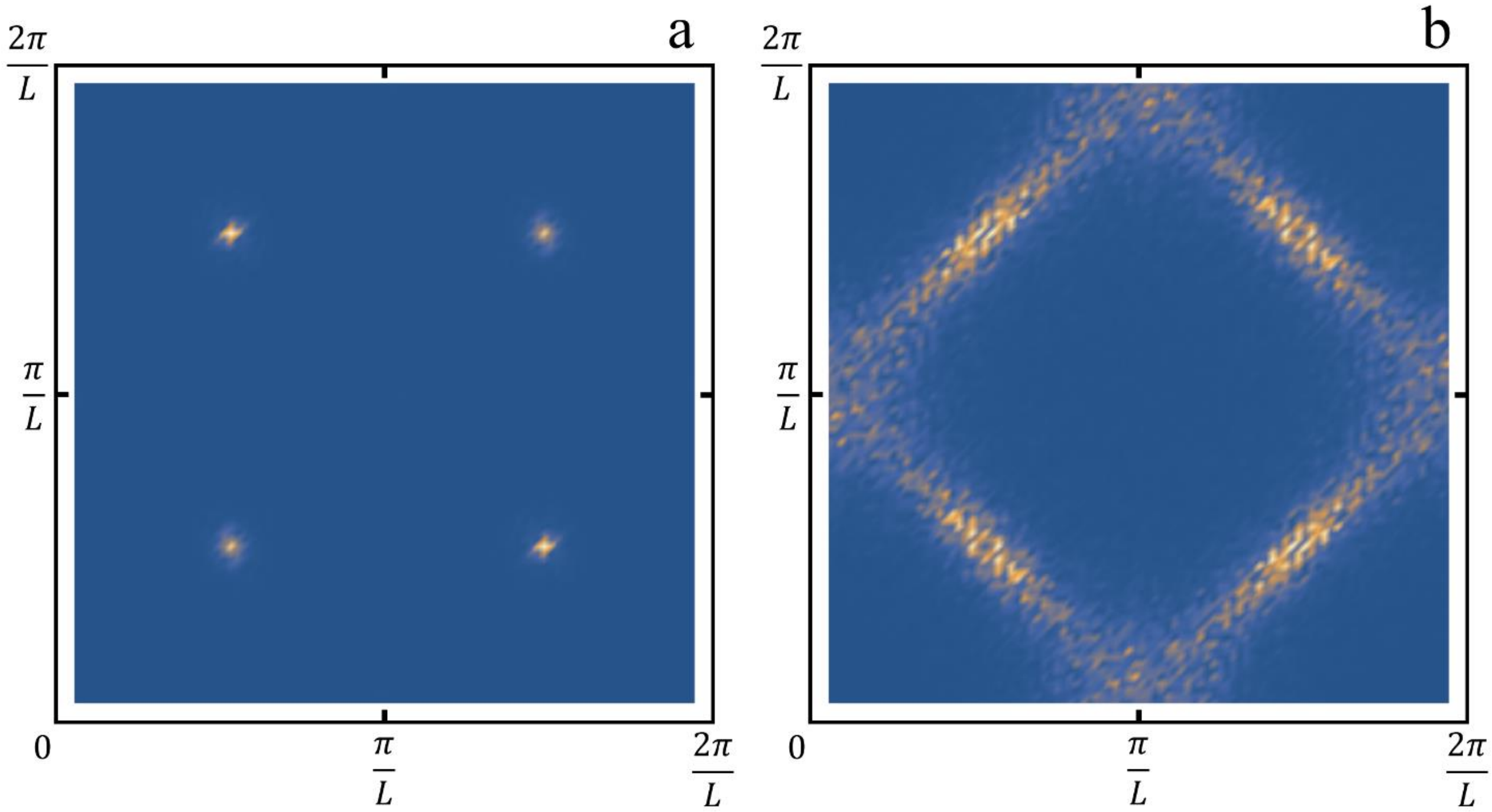

**Figure 2.** Reciprocal lattice image at T = 0.001; $J_2$ = 0.3 (a), $J_2$ = 3 (b), other parameters are fixed: $U$ = 1, $V$ = 1, $t$ = 3, $J_1$ = 3, $J_3$ = 1.5.

Despite the fact that the nearest-neighbor exchange integral exceeds the others, AFM nearest-neighbor ordering and the corresponding structure factor value are not observed due to charge disproportionation, since the nearest neighbors of the spin-triplet $Ni^{2+}$ centers become nonmagnetic $Ni^{4+}$ centers. Thus, the presence of charge order suppresses this AFM phase (with $\vec{k}$ = (1/2, 1/2)). Consequently, in such a system, magnetic order can arise only due to the interaction of the second neighbors: $Ni^{2+}$-$Ni^{2+}$, located at an angle of 90°, and the third: $Ni^{2+}$-$Ni^{2+}$ at an angle of 180°. In Fig. 1(a), AFM ordering of the third neighbors is observed in the system. This magnetic structure has a propagation vector $\vec{k}$ = (1/4, 1/4) and it's a two-dimensional analogue of the structure with $\vec{k}$ = (1/2, 0, 1/2) or $\vec{k}$ = (1/4, 1/4, 1/4) in pseudocubic notation, which is consistent with experimental data. The appearance of such magnetic structure is possible only in the presence of charge ordering. The Hamiltonian parameters for finding the magnetic structure, in particular the ratio of the exchange integrals, were chosen based on the geometry of the model nickelate and the exchange integrals calculated from first-principles calculations [12]. Increasing the exchange interaction of the second neighbors will gradually blur the structure with $\vec{k}$ = (1/4, 1/4) (Fig. 1.b), and at a sufficiently large $J_2$, a structure $\vec{k}$ = (1/2, 0) will form. This structure is also formed only in the presence of charge order. Figure 3 shows the spin sublattice for the situations depicted in figure 1 and demonstrate the magnetic ordering of the third neighbors (a) and the competition between the magnetic orders of the second and third neighbors (b).

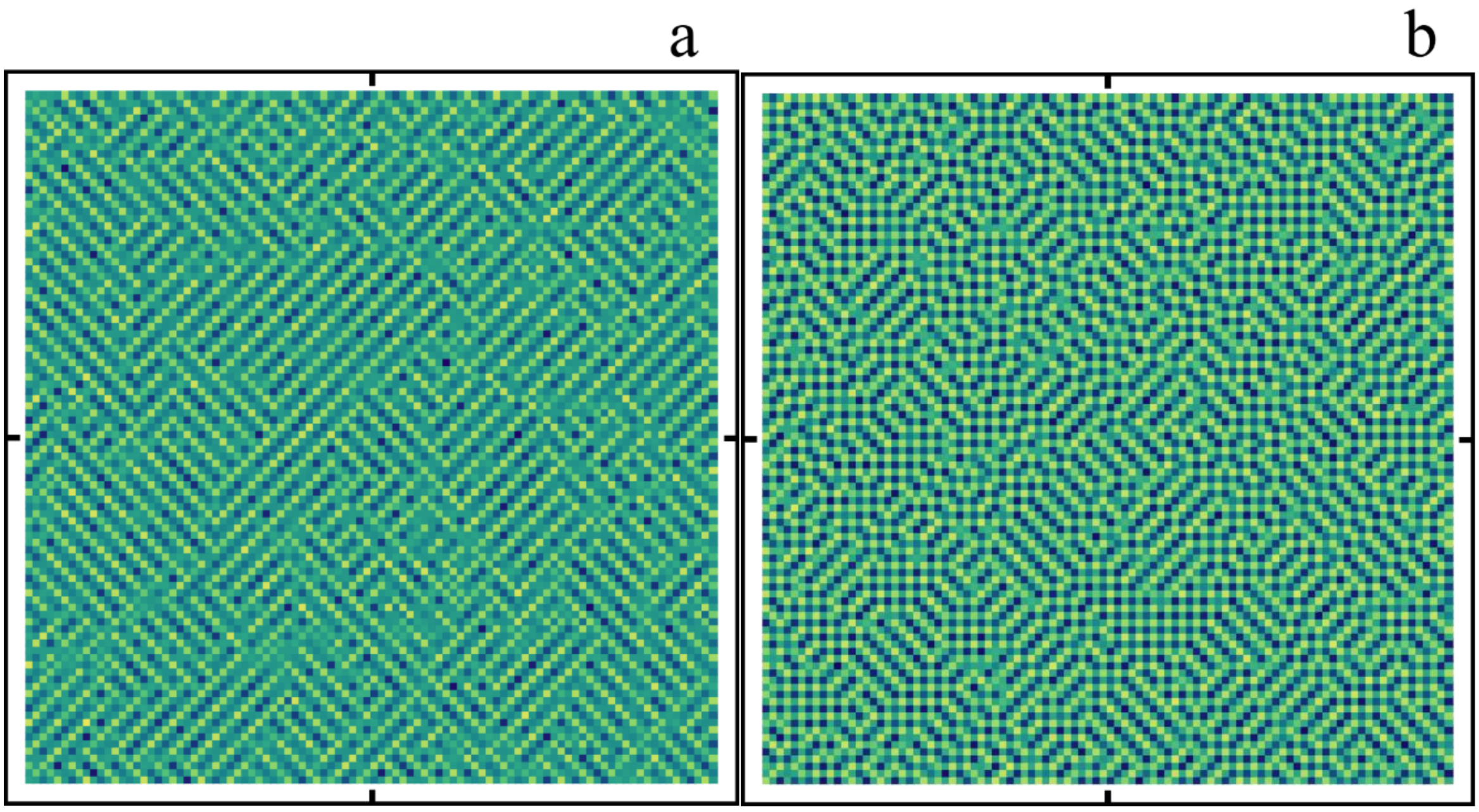

**Figure 3.** Spin sublattice with magnetic ordering of third neighbors (a) and competition of magnetic orders of second and third neighbors (b); yellow – $\sigma = +1$, green – $\sigma = 0$, blue $\sigma = -1$.

## CONCLUSIONS

Using the classical Monte Carlo method with kinematic account of constant bosons concentration a two-dimensional model nickelate was studied, its magnetic structures for various values of the exchange integrals of the first, second, and third neighbors were determined. It was shown that, for a certain ratio of the exchange integrals and the presence of charge ordering in the system, a magnetic structure with a propagation vector $\vec{k}$ = (1/4, 1/4) is formed, which is a two-dimensional analogue of the phase with $\vec{k}$ = (1/2, 0, 1/2) or $\vec{k}$ = (1/4, 1/4, 1/4) in pseudocubic notation which is consistent with experimental data.

**Funding.** This work was supported by the Ministry of Science and Higher Education of the Russian Federation, project FEUZ-2026-0014.